\pdfoutput=1
\documentclass{JINST}

\title{The design and basic performance of a Spiral Fiber Tracker for the J-PARC E36 experiment }

\author{O.~Mineev$^a$\thanks{Corresponding author.}, S.~Bianchin$^b$, M.~D.~Hasinoff$^c$, K.~Horie$^d$, Y.~Igarashi$^e$, J.~Imazato$^e$, H.~Ito$^f$, H.~Kawai$^f$, S.~Kodama$^f$, M.~Kohl$^i$,  Yu.~Kudenko$^{a, j, k}$, S.~Shimizu$^d$, M.~Tabata$^f$, A.~Toyoda$^e$,  N.~Yershov$^a$\\
\llap{$^a$}Institute for Nuclear Research of RAS, Moscow, Russia\\
\llap{$^b$}Canada's National Laboratory for Particle and Nuclear Physics (TRIUMF), Canada\\
\llap{$^c$}University of British Columbia, Canada\\
\llap{$^d$}Osaka University, Japan\\
\llap{$^e$}High Energy Accelerator Research Organization (KEK), Japan\\
\llap{$^f$}Chiba University, Japan\\
\llap{$^i$}Hampton University, USA\\
\llap{$^j$}National Research Nuclear University MEPhI, Moscow, Russia\\
\llap{$^k$}Moscow Institute of Physics and Technology, Russia\\
 E-mail: \email{oleg@inr.ru}}

\abstract{A spiral fiber tracker (SFT) has been designed and produced for the J-PARC E36 experiment as  an element of the tracking system for conducting a high-resolution momentum measurement of charge particles from kaon decays. A novel technique to wind the pre-made fiber ribbons spirally was employed for the configuration with four detector layers made of 1~mm diameter plastic scintillating fibers. Good position alignment and sufficiently high detection efficiency for charged particles with minimum ionizing energy were confirmed in cosmic ray test. The tracker was successfully used in the E36 experiment.}  


\begin{document}
\section{Introduction}

The TREK/E36 experiment \cite{e36} has been performed at the Japan Proton Accelerator Research Complex (J-PARC) \cite{jparc}. The primary physics goal of the experiment is the search for lepton universality violation in the precise measurement of the decay width ratio  $\Gamma (K^+\rightarrow e^+\nu) /\Gamma (K^+\rightarrow \mu^+ \nu)$ using stopped positive kaons. Since the Standard Model (SM) prediction for the ratio is highly precise,  a deviation would clearly indicate the existence of New Physics beyond the SM. 

\begin{figure}[htb]
\begin{center}
\includegraphics[width=13cm,angle=0]{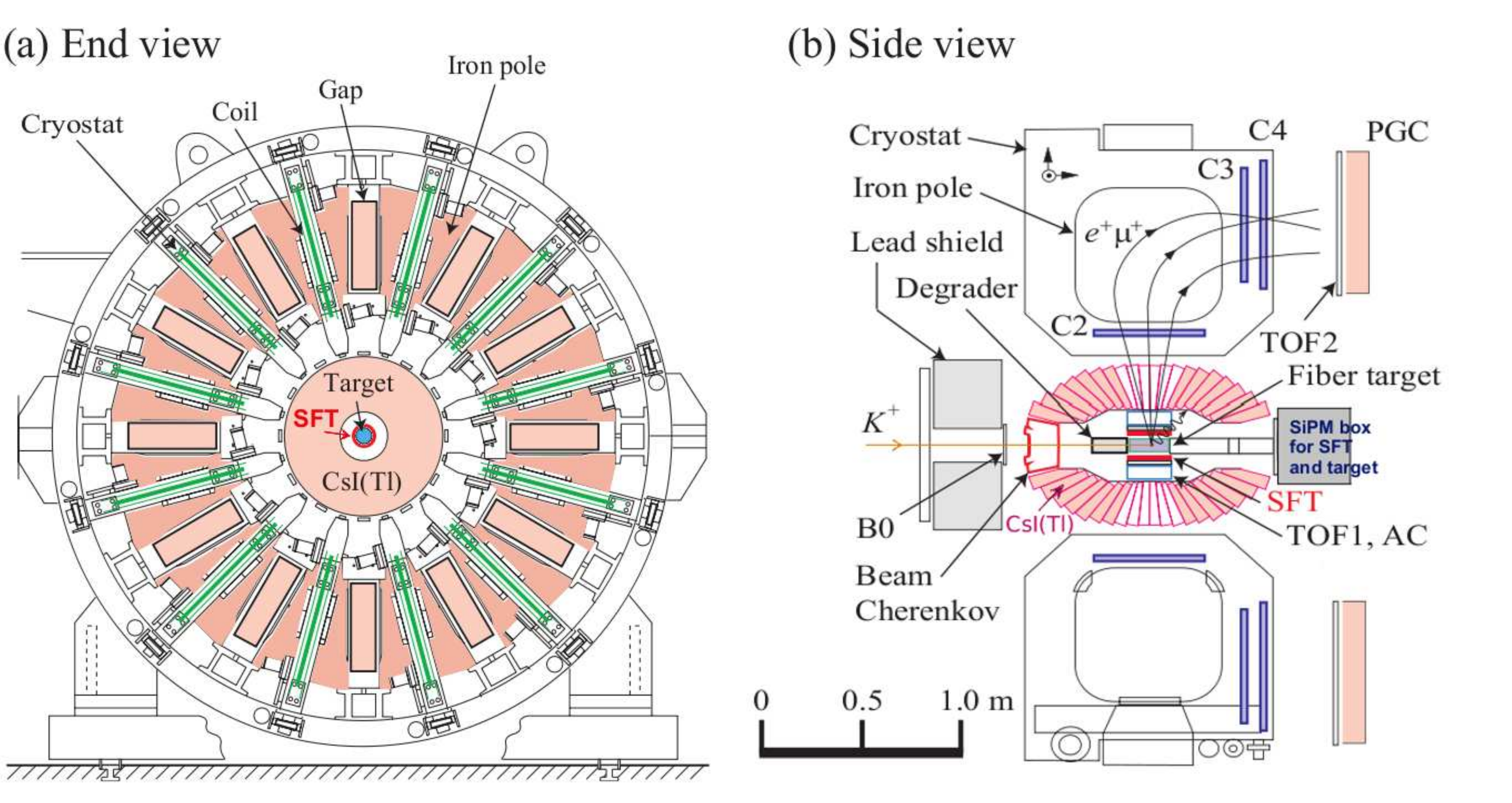}
\caption{End (a) and cross-sectional side (b) views of the E36 detector setup.}
\end{center}
\label{fig:detector} 
\end{figure}

Schematic cross-sectional side- and end views of the E36 detector are shown in Fig.~\ref{fig:detector}. The setup represents an upgraded version  of the apparatus for the previous  E246 experiment at the KEK 12-GeV proton synchrotron \cite{e246}.
The detector comprises a $K^+$ stopping fiber target, a charged particle tracking system, a particle identification (PID) system and a CsI(Tl) photon detector, in conjunction with a 12 gap super-conducting toroidal spectrometer of 1.4~T central magnetic field. 
PID between $e^+$ and $\mu^+$ is carried out with aerogel Cherenkov counters (AC), lead-glass Cherenkov counters (PGC) and time-of-flight measurement with TOF1 and TOF2 counters. 

The tracking system determines the trajectory of each $K^+$ decay product with 3 layers of multi-wire proportional chambers (C2, C3 and C4) in each magnet gap  along with the decay vertex information in the central region. This provides a measurement of the momentum in the interesting region up to about 300 MeV/$c$\footnote[1]{Electrons and muons in the relevant decays have the momentum of 247~MeV/$c$ and 236~MeV/$c$, respectively.}. The target as a vertex detector is a bundle of 256 pieces of 3~mm rectangular scintillating bars with an overall diameter of 68~mm and 200~mm length. The target provides measurement of the $x$ and $y$ coordinates of the decay vertex normal to the beam and also the azimuthal angle $\phi$ of the outgoing secondary particles. The $z$ coordinate along the beam axis with the resolution of 1~mm is given by the spiral fiber tracker (hereafter SFT) described in this report. Since the SFT and the target  both have light readout via plastic fibers  we use the same photosensors and readout electronics for both detectors.

\section {Design of the SFT}

\subsection{Concept of the SFT} 

The tracking detector for the $z$-coordinate has to be accommodated inside of the existing CsI(Tl) detector together with the target, TOF1 and AC. The diameter of the CsI(Tl) aperture was only 225~mm. Because all the elements require thickness as large as possible, the space available for the SFT detector is limited to be less than 5~mm in thickness.  The minimization of the detector cost is also an important consideration. 

A novel technique was employed to spiral the pre-made scintillating fiber ribbons around a  cylindrical bobbin  and to read out light from both ends of the ribbons. For a given bobbin diameter, the fiber stereo angle is determined by the width of the ribbons, namely by the number of fibers in a ribbon. As long as the stereo angle is not large the $z$-detection resolution is essentially determined by the fiber diameter. 

\subsection{Details of the design}

In designing the SFT the following conditions were considered:  1) Several layers of the fiber ribbons are necessary to ensure the high detection efficiency and to extract the hit position information. 2) It was essential to wind the fiber ribbons with two different helicities with positive $\theta_{L}$ and  negative $\theta_{R}$ in order to obtain crossing point. 3) The length of the ribbons should be minimized because of the light attenuation in the fibers. 4) The ribbons must be flexible enough for winding around the bobbin with a diameter of 79~mm, which serves as the support housing of the target, and to route the ribbon ends through the central setup region to the photosensor box. 5) The number of readout channels must be minimized. The number is determined by the ribbon width once the fiber diameter has been fixed. 

\begin{figure}[htb]
\begin{center}
\includegraphics[width=13cm,angle=0]{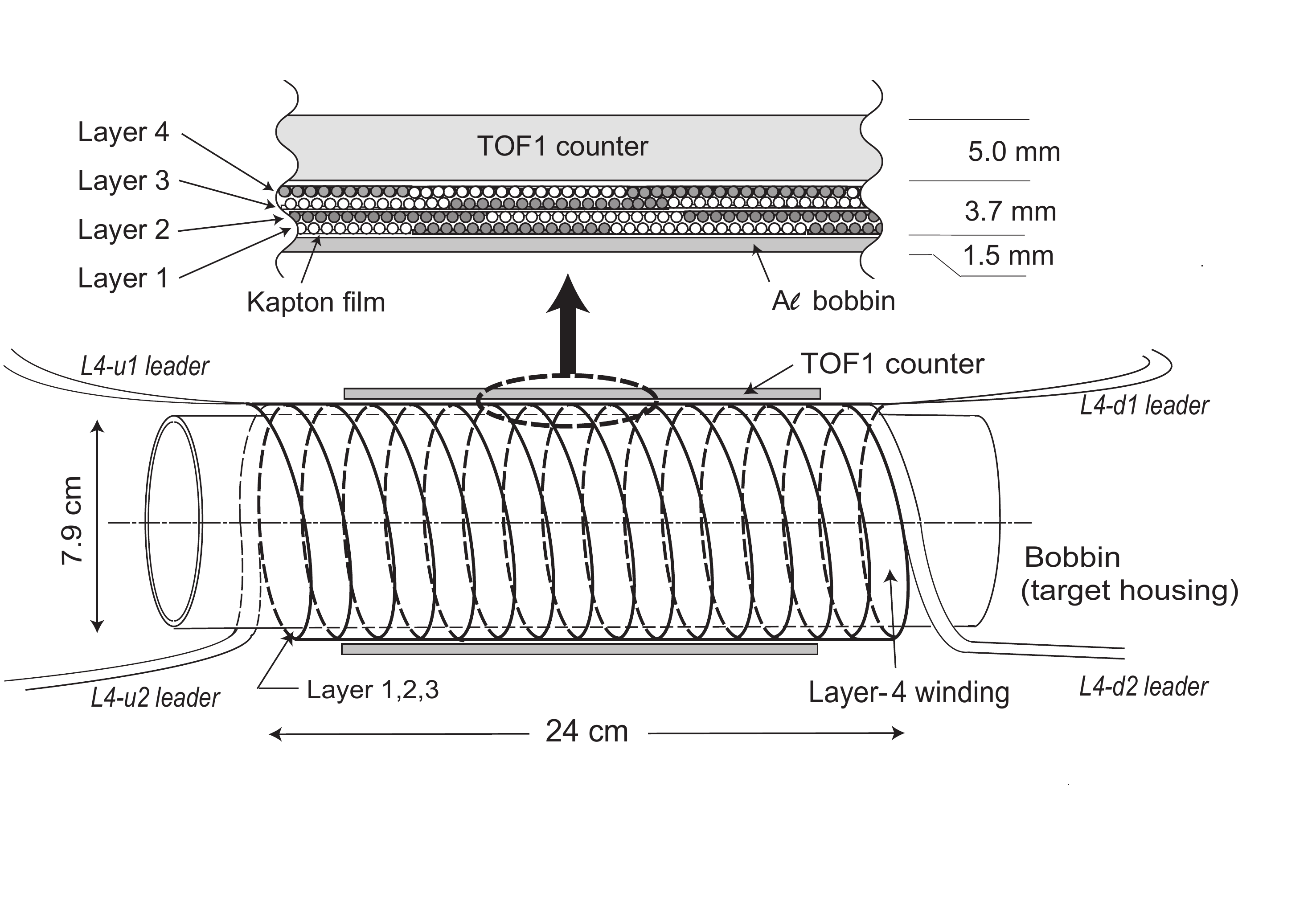}
\caption{Schematic view of the SFT. Only the winding of the 4th layer is shown for
simplicity. The cross section of the four coiled ribbon layers is shown enlarged, where open and filled circles indicate neighboring turns of the same ribbon.  The ribbons are separated at each end into two bundles for routing the fibers to the photosensors.}
\label{fig:sft_view} 
\end{center}
\end{figure}

It was essential to take different $\vert\theta_{L}\vert$ and $\vert\theta_{R}\vert$ in order to resolve the degeneracy of azimuthal angle $\phi$ determination as will be discussed below. Thus, different ribbon widths  must be adopted for the left-helicity  and right-helicity ribbons. After simulating several different combinations for the  width, the number of 1~mm fibers in the ribbons  was adopted as $N_L=15$ and $N_R=17$. The 1-mm fiber diameter was selected after considering the detection efficiency expected and the  condition  4). This choice enabled the structure with two $L$-helicity layers and two $R$-helicity layers, also satisfying the conditions 3) and 5). 
As shown in the schematic picture in Fig.~\ref{fig:sft_view}, the fourth layer of 14 turns covers the 20~cm target.
 
\subsection{Principle of hit position reconstruction}

\begin{figure}[htb] \begin{center} 
\includegraphics[width=11cm,angle=0]{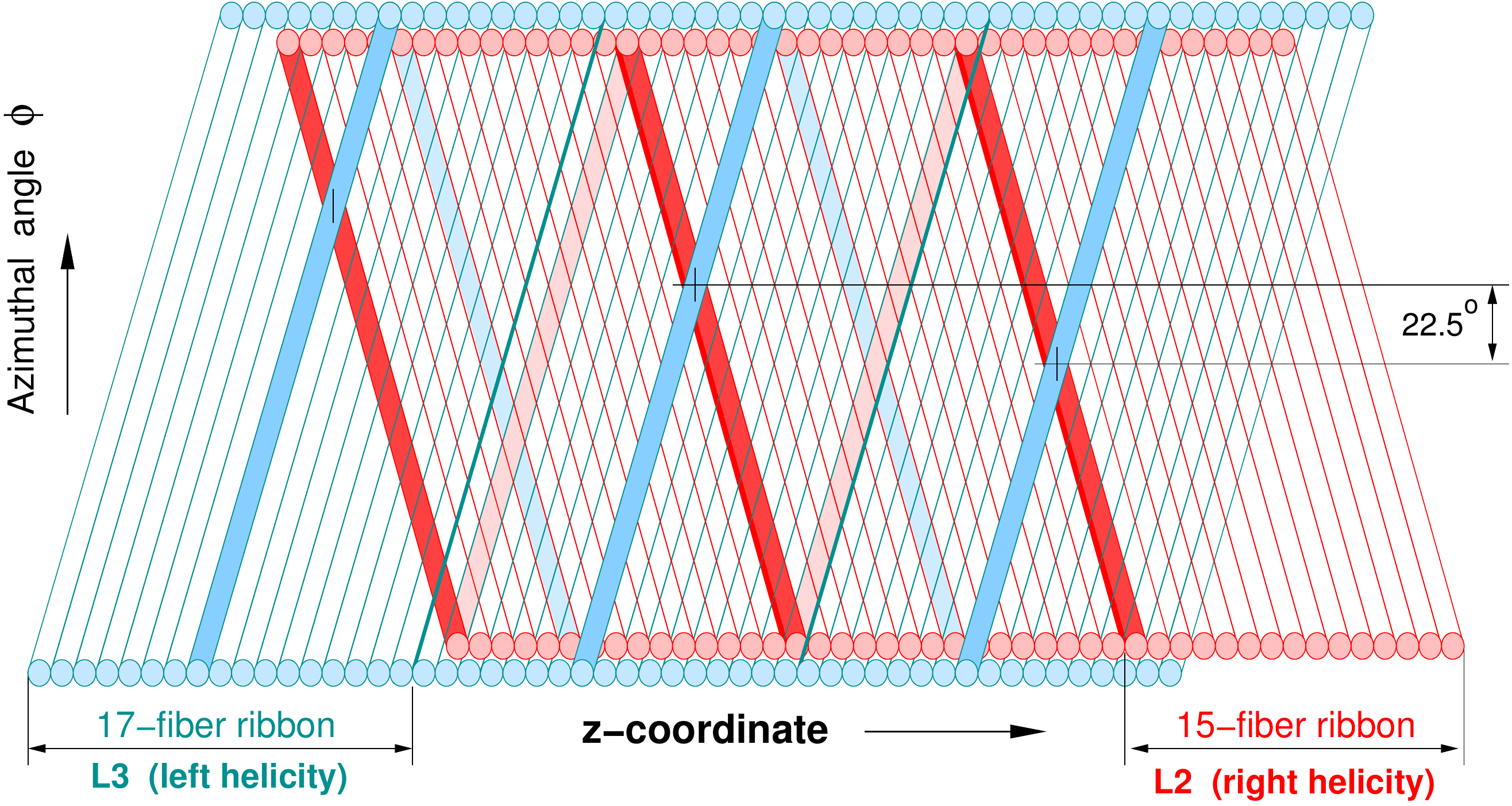}
\end{center}
\vspace{0cm}
\caption{Development of the SFT coil schematically drawn with the exaggerated stereo angle. The $L$-helicity ribbon with 17 fibers (L3 layer) and $R$-helicity ribbon with 15 fibers (L2 layer) are indicated in blue and red, respectively. The azimuthal angle $\phi$ of crossing point of the hit fibers changes by $22.5^{\circ}$ from turn to
turn.}
\label{fig:helicity} 
\end{figure}

The development of the SFT winding is schematically illustrated in Fig.~\ref{fig:helicity} showing a single layer from each helicity with 17-fiber $L$-helicity  and  15-fiber $R$-helicity ribbons. A charged particle passage produces a signal in both the $L$- and $R$-ribbons. The relevant hit fibers are indicated with a blue strip for the $L$-helicity and a red one for the $R$-helicity. However, it is not known in which turn of the ribbon the hit occurs. There are several possible candidate cross points indicated in Fig.~\ref{fig:helicity} with the azimuthal angle steps of 22.5$^{\circ}$. By using the azimuthal angle information from the target $x/y$ data, the real hit point can be uniquely determined thus providing the exact $z$-coordinate with an accuracy equal to the fiber diameter.   

\section{Manufacturing of the SFT}

\subsection{Test of the scintillating fibers}

We have chosen plastic scintillating fibers Kuraray SCSF-78MJ~\cite{fiber} of 1~mm diameter because they are intrinsically fast and have a high degree of geometrical adaptability.  Their peak emission wavelength at 450~nm matches well the peak sensitivity of the employed photosensor, Hamamatsu MPPC S10362-11-50C \cite{mppc}. Table~\ref{table:fiber} lists Kuraray specification for the fiber. 
%
\begin{table}[h]
\caption{ Specification of Kuraray SCSF-78MJ fiber \cite{fiber}}
 \begin{center}
 \begin{tabular}{|l|c|l|}  
 \hline
 Fiber type  &  Kuraray SCSF-78MJ & $\bullet$ multi-clad, none-S type \\
Diameter        & 1.0~mm    & $\bullet$ tolerance $\pm$10~$\mu$m  \\
Cladding thickness & 60 $\mu$m & $\bullet$ two layers of cladding  \\
Decay time  & 2.8 ns  &  \\
Emission peak & 450 nm (blue) & $\bullet$ full spectrum within 420--550 nm   \\
Attenuation length & > 4 m    & $\bullet$ 3~m long fiber read out by PMT\\
Bending loss of light  & 3--4~\%  & $\bullet$ per a full circle with D=7~cm \\
Minimum bending diameter & 4--5~cm & $\bullet$ the rapid increase of the light loss \\
             &         & due to cracking of core  \\
\hline
\end{tabular}
\end{center} 
\label{table:fiber}  
\end{table}

Although there are several data on this fiber in the literature (\cite{gluex1}, \cite{gluex2}), a few 3~m long samples were tested to confirm the basic expected performance of the SFT. The light output measurements were done with a $^{90}$Sr $\beta$-ray source. The source irradiated the fiber through a 1.4~mm diameter lead collimator. The trigger signal was produced by a $1\times1$~cm$^2$ plastic scintillator counter located behind the fiber. This counter was viewed by a small photomultiplier. The fiber readout was done with an MPPC  of the same type adopted for SFT. The other fiber end was left open without a mirror.

The average light yield was measured to be 5--6~$p.e.$ at a distance of 3~m from the MPPC corresponding roughly to the center of SFT.  No correction for optical crosstalk of the MPPC was applied, so the result can be overestimated by 20~\%. Taking into account the bending loss  for 8 turns (corresponding roughly to the center of the SFT) estimated from Kuraray specification and the anticipated junction loss from the scintillating fiber to the clear fiber, the light yield was estimated to be around 3~$p.e.$ from a single end for the central part of the SFT. A time resolution $\sigma_t$=2.0~ns was also obtained at this distance.

\begin{figure}[htb] \begin{center} 
\includegraphics[width=10.0 cm,angle=0]{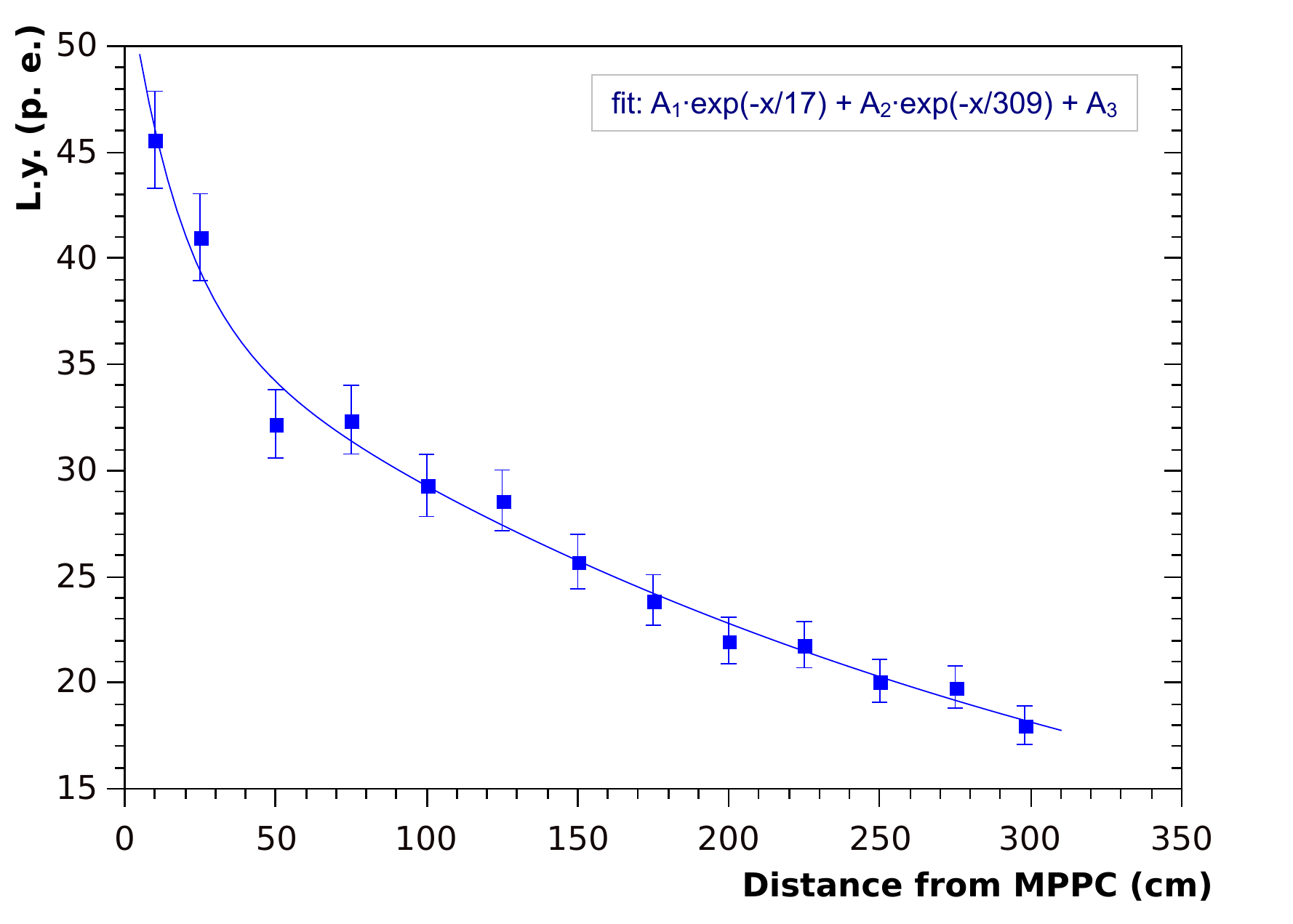}
\end{center}
\vspace{0cm}
\caption{Light attenuation curve of SCSF-78MJ measured with a  1~mm diameter 3~m long sample fiber. The light yield (l. y.) in number of photoelectrons ($p. e.$) was measured as a function of the light source distance $d$ from the MPPC. A double exponential function was used to fit the data. }
\label{fig:atten} \end{figure}

The light attenuation was measured by illuminating the fiber with a UV LED with a peak emission wavelength of around 365~nm. The LED excited the fiber through a collimator of 0.5~mm diameter at a variable distance $d$ from the MPPC. The other end of the fiber was optically open. The results are shown in Fig.~\ref{fig:atten}. The data were fitted by a sum of two exponential functions: 
$A_1\cdot$~exp$(-d/\lambda_s)+A_2\cdot$~exp$(-d/\lambda_l)+A_3$, where the short attenuation length $\lambda_s$=17~cm and the longer attenuation length $\lambda_l$=309~cm. The latter value was relevant.

\subsection{Structure of the ribbons }

\begin{figure}[htb] \begin{center} 
\includegraphics[width=12cm,angle=0]{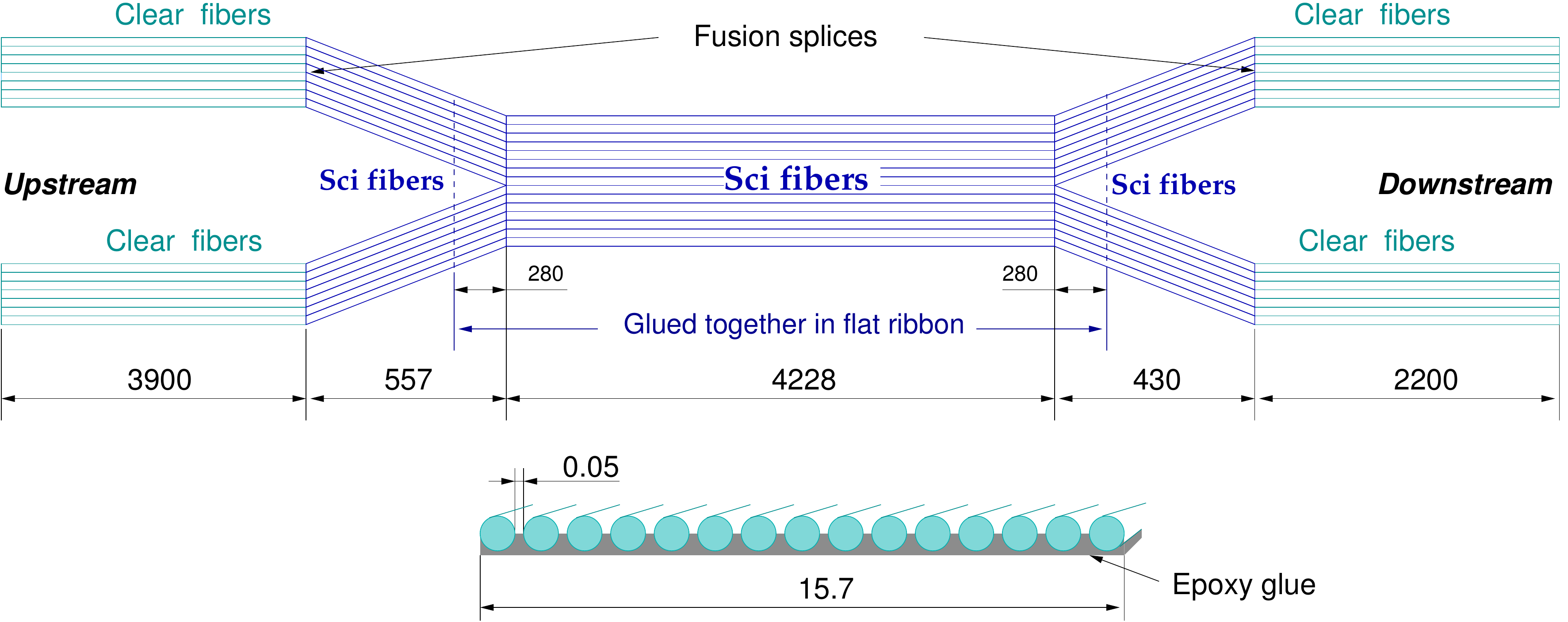}
\end{center}
\vspace{0cm}
\caption{Schematic layout of the ribbon for 1st layer. All dimensions are in mm.}
\label{fig:ribbon} \end{figure}

The four fiber ribbons have slightly different lengths because of the slightly different layer diameters. A schematic drawing of one of the ribbons is shown in Fig.~\ref{fig:ribbon}. The 5~m long active scintillating (SciFi) part is spliced at both ends with a 1~mm clear Kuraray fibers (CF)  to route the light signals downstream outside the magnet to the photosensors. The splicing  was necessary in order to reduce the light attenuation and to avoid the noise of charged particles passing through the light-guide part. The length of the SciFi was minimized while ensuring the sufficient length of the active part. 
The CF length is 3.9~m and 2.2~m at upstream and downstream side, respectively, forming a whole fiber with the total length of 11.5~m. The ribbon is split at both ends in two narrower ribbons to facilitate the fiber routing of the SFT active cylindrical part. The end part of the SciFi and the CF are  not glued to a ribbon and they form a flexible bundle. 
The individual fiber length and ribbon size  for each  layer are presented in Table~\ref{table:layers}.

\begin{table}[h]
\caption{Parameters of the SFT layers.}
 \begin{center}
 \begin{tabular}{|l|c|c|c|c|}  
 \hline
~~~~~~~~~~Parameters  &  Layer 1 & Layer 2 & Layer 3 & Layer 4\\
\hline
Winding helicity  & right  & right  & left  &  left \\
SciFi length (cm)  &  521.5   &  530.7    &  462.3     &  470.2  \\
CF length (cm)     &  220, 390 &  220, 390  & 220, 390    & 220, 390   \\
Number of fibers       &   15     &  15       &  17        &   17    \\
Ribbon width (mm)  &   15.75  &  15.75    &  17.85     &   17.85 \\
Diameter of layer (mm)      &   80.5   &  82.2     &  84.2      &  85.9   \\
Number of ribbon turns &   17     &  17       &   14       &  14   \\
\hline
\end{tabular}
\end{center} 
\label{table:layers}  
\end{table}

\subsection{Production of the fiber ribbons}

After confirming the satisfactory charactersitics of the sample fiber, Kuraray SCSF-78 fibers with 1.0~mm diameter were adopted for the ribbon production.  The basic performance of a single  and also a double  layer of the staggered ribbons was investigated using a prototype ribbon to confirm the acceptable detection efficiency \cite{tabata}. 
Two ribbons with 15 fibers for layers L1 and L2, and two with 17 ribbons for layers L3 and L4 were manufactured\footnote{The production was ordered to Moderation-Line Co., Ltd (Japan) which had previous experiences in fiber ribbons production and fusion of plastic fibers.}. 
The issues during the ribbon production were:  1) how to retain the integrity of a ribbon during the winding while keeping the bending flexibility; 2) how to ensure the positioning precision of each fiber within the ribbon, and 3) the optical characteristics of the fibers in the ribbon such as additional light attenuation and crosstalk were also investigated. 

The gluing was done on a metal jig with grooves precisely machined with the interval of 1.050~mm. Since the fiber diameter can deviate $\pm$10~$\mu$m, the space between neighboring fibers was set to be about 50~$\mu$m to fix an accurate position of each fiber in the ribbon. Several types of glues were tested and a soft epoxy was adopted satisfying the conditions above. The epoxy glue was applied only to one side of the ribbon as shown in Fig.~\ref{fig:ribbon}. The repulsive force (bending rigidity) for a sample ribbon with 16 fiber glued was measured to be about 1.29~N at a bending radius of 4~cm showing only 20~\% contribution from the glue. No significant additional attenuation and crosstalk were observed after gluing. 

\begin{figure}[htb] \begin{center} 
\includegraphics[width=9cm,angle=0]{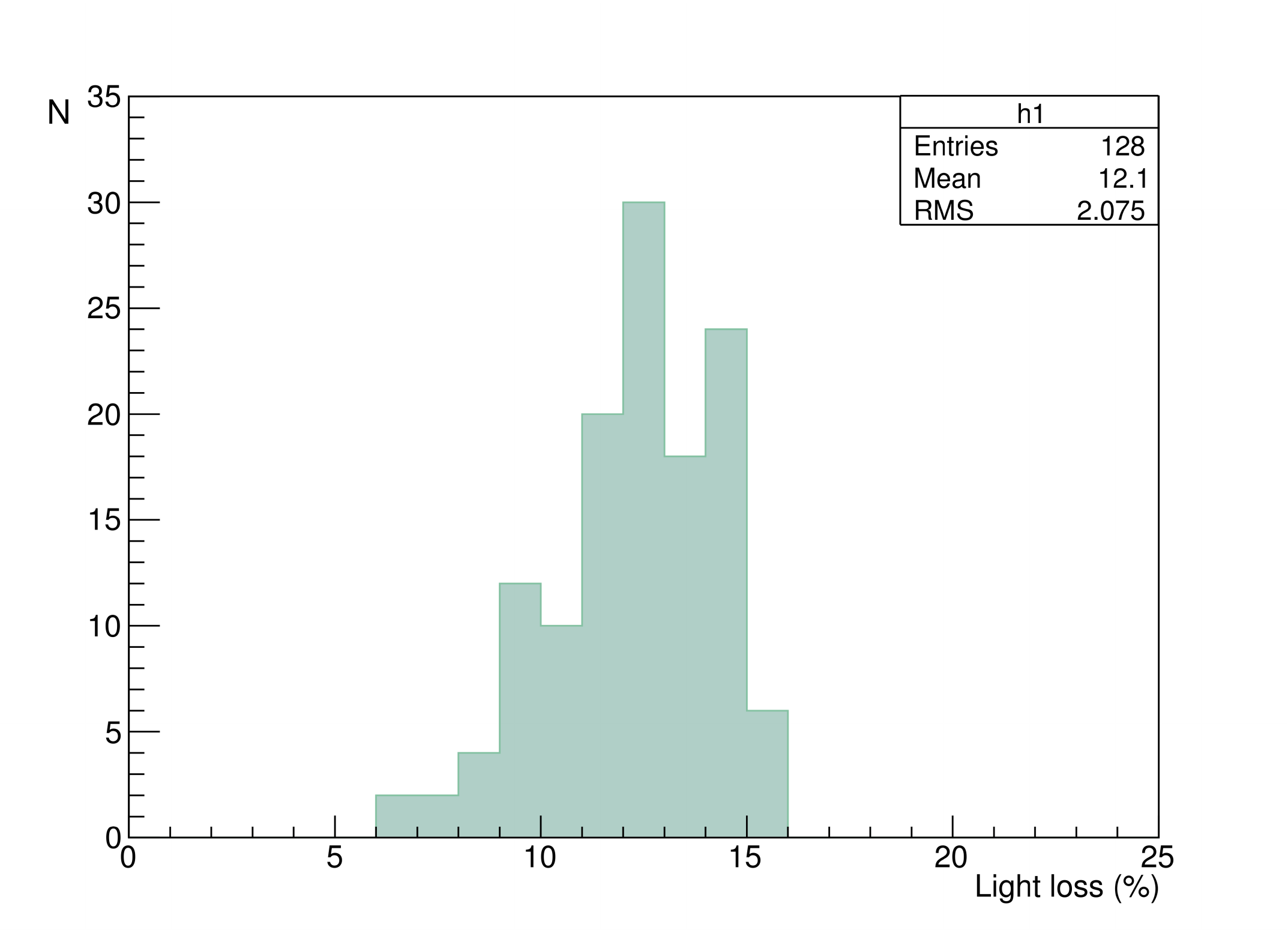}
\end{center}
\vspace{0cm}
\caption{Distribution of light losses in 128 joints between the SciFi and CF fibers. }
\label{fig:ll} \end{figure}

After the ribbon gluing the SciFi and CF fibers were spliced by heat-fusing one by one and the junction part was covered with a 2~cm long Teflon sleeve for mechanical protection. The loss at all the 128 junctions was measured with a power meter. The distribution of the loss-per-joint is shown in Fig.~\ref{fig:ll}. Assuming that the losses were equal in both joints of a fiber, the average loss per joint was around 12~\% and the maximum light loss was smaller than 16~\%.

\subsection{Assembling }

A rotating dummy pipe with the same diameter as the actual bobbin (target holder) was prepared for the SFT assembling. A Kapton film of 0.15~mm thickness was wrapped around the pipe  by one turn as a smooth substrate for sliding the SFT during the installation process. To facilitate the removal of the winding from the pipe, eight strips of 0.1~mm thick Mylar film were placed along the axis between the pipe and the Kapton film. The fiber ribbons were tightly coiled around the dummy pipe and fixed locally by  adhesive tape. Although a bending diameter of larger than 10 cm was the Kuraray recommendation for long term application, a smaller radius was thought to be tolerable for our short-term use. Actually the visual inspection did not reveal any damage in the fiber cladding. The layers 1 and 3 were wound with the glue surface inside and the layers 2 and 4 opposite to make staggering between 1~\&~2 and 3~\&~4, respectively. The total thickness of the four layers amounts to 3.7~mm (less than 4~mm due to staggering).

\begin{figure}[htb] \begin{center}
\includegraphics[width=15.5cm,angle=0]{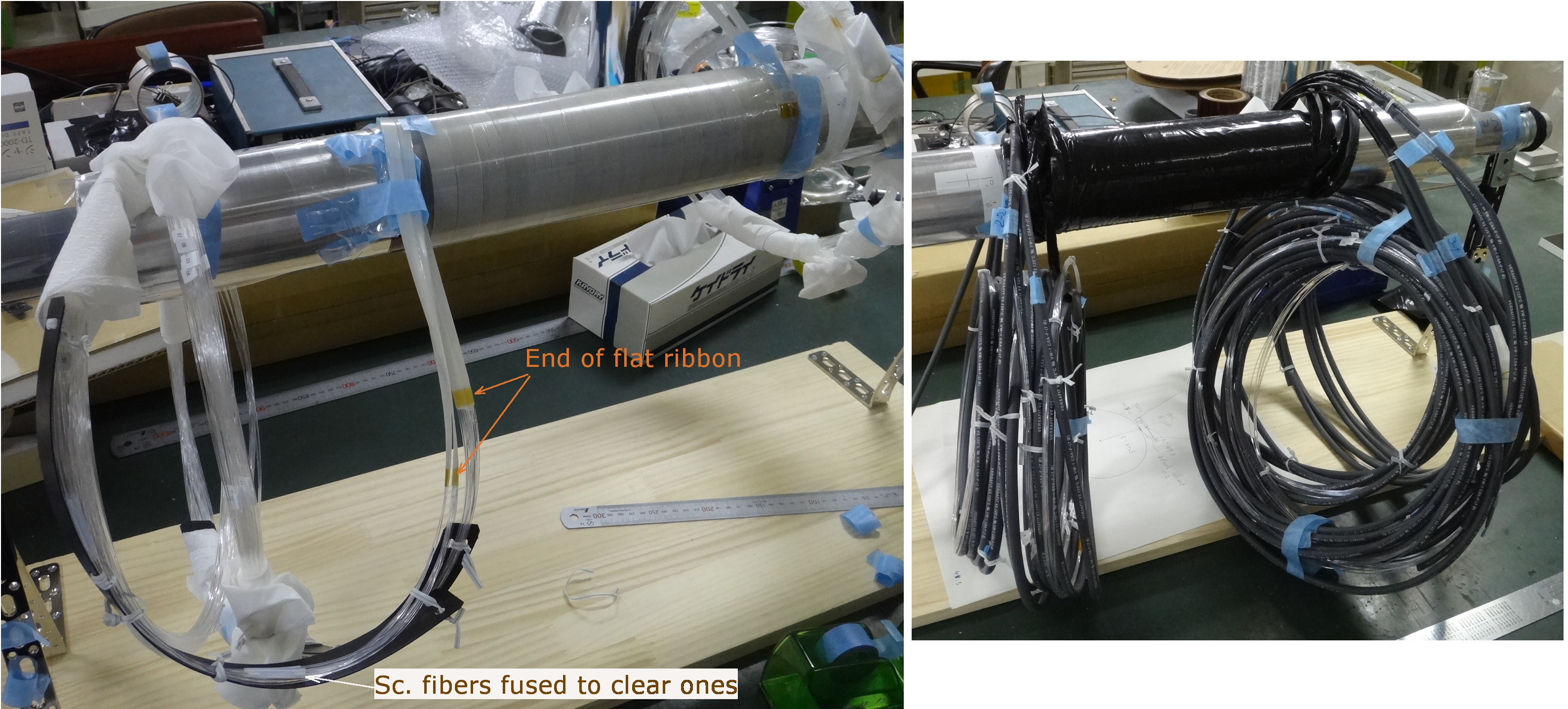}
\end{center}
\vspace{0cm}
\caption{Assembling of the SFT: first two ribbons are coiled around the dummy pipe ({\it left}); SFT view before gluing of the fibers into couplers ({\it right}).}
\label{fig:assembly} 
\end{figure}

The active part of the SFT of the 4th layer represents 14 turns of the 17-fiber ribbon, and thus there are 238 fibers in total in the fiducial region which is about 240~mm in length. The outer diameter is 86~mm.  Fig.~\ref{fig:assembly} shows pictures at different assembly stages.
The coiled part of the ribbons and the outlet CF leads were shielded from light with black plastic sheet and flexible black plastic tubes. The ends of clear fibers were then glued into  plastic MPPC couplers \cite{coupler}  with colorless epoxy cement, and the fiber ends were hand-polished. Finally, the ribbon winding was transferred from the dummy pipe onto the actual bobbin (target holder). All the fibers were examined with an LED flasher to check for any accidental damage inflicted during assembly. All assembly procedures were carefully performed to avoid applying extra bending tension in the fiber ribbons and bundles. 

\section{Performance check}

\subsection{Initial bench test with a radioactive source}

The completed SFT was tested with a $\beta$-ray source. The newly developed EASIROC board (next subsection) was commissioned at the same time. The signal characteristics as functions of the MPPC temperature and the bias voltage were investigated. Then, the position dependence of the pulse height distribution was observed using a collimator placed between the SFT and the source placed inside the 1.5~mm thick Al-bobbin.

The relative positions of the fibers in Layers 1/2 and  Layers 3/4 were checked. The staggering pattern was confirmed to be perfect for the L1 and L2, however, some disordering (two fibers maximum) depending on the azimuthal angle  was observed for  L3 and L4, which can be corrected in the off-line analysis.

\subsection{Photosensors and front-end electronics}

A total of 64 fibers were read out at both ends by 128 Hamamatsu MPPCs S10362-11-50C. The MPPCs have an active area of $1\times1$~mm$^2$, 400 pixels, and the photon detection efficiency for a blue scintillating light is above 30~\% after correction for crosstalk and after-pulsing. The MPPCs were inserted in couplers and soldered onto the 8 channel carrier boards which  mounted inside a light-tight box cooled by Peltier elements to $5-8^{\circ}$C. Signals from the MPPCs were sent directly to VME ADCs over 9.2~m long 8-channel micro coaxial ribbon cables. 

The MPPC readout is implemented with 64-ch. VME-EASIROC boards designed by the Tohoku University group \cite{easiroc}. The design is based on the ASIC EASIROC (Extended Analogue Sipm Integrated Read-Out Chip) developed by Omega/IN2P3 \cite{omega}. The board combines an amplifier, peak-hold ADC with low and high gains, a discriminator, a multi-hit TDC with 0.8~ns resolution, and a bias voltage adjustment for each individual MPPC. 
The 74~V bias voltage is supplied through standard coaxial cables to the carrier boards, and the fine adjustment to the bias voltage reaches the MPPCs through the  micro coaxial cables.

\subsection{Detection efficiency in a cosmic ray test}

Cosmic ray tests were performed to measure the SFT detection efficiency with the actual cables and electronics to be used in the experiment. The trigger was arranged with the TOF1 counters plus an additional trigger counter which limited   the hit area to $\Delta z$=5~cm. Only cosmic muons incident normal to the SFT surface were selected.  Because of the long cables and induced noise, we could not resolve the individual photoelectrons in the ADC spectrum. The TDC data were used for event selection.

Cosmic ray particles cross the four fiber layers in the upper part of the SFT and the four fiber layers in the lower part generating two hit signals for each trigger, except for the case where the penetrating particle hits the same fiber in both the upper and lower parts in a ribbon. This effect was taken into account in the  total counting  statistics.
Because of the low light yield an event was counted if either of the fiber ends fired, i.e. contained a hit within the timing peak of the TDC spectrum. 
Contributions from backgrounds and accidentals in the total number of detected events were calculated by analyzing the raw TDC spectra. 

The detection efficiency for a fiber was calculated as the ratio of selected events  to the number os triggers,  and that of a layer was just the sum of the fiber efficiencies. The results are listed in Table~\ref{table:eff}. The measured value of the layer efficiency contains the contribution from geometrical inefficiency caused by the 50~$\mu$m gap between the scintillating fibers plus the inactive fiber cladding of 60~$\mu$m thickness, which amounts 16~\% in total. The third column in Table~\ref{table:eff} shows the corrected values of the layer detection efficiency with this geometrical inefficiency removed. The geometrical inefficiency can be suppressed  due to fiber staggering configuration between L1/2 and L3/4.

The inefficiency without any hits in the four layers was measured to be 0.17~\% without correction for accidentals. The time resolution  $\sigma_t$ for a single fiber was found to be around 3.0~ns for the average between times from the two fiber ends ($t_1+t_2$)/2. 
\begin{table}[h]
\caption{Detection efficiency for the SFT layers. The third column shows the detection efficiencies after correction for the geometrical dead space between the fibers.}
 \begin{center}
 \begin{tabular}{|c|cc|}  
 \hline
  Layer  & Efficiency (\%) & Corrected (\%) \\
 \hline
 1   & 77.1 & 91.8  \\
 2   & 79.1 & 94.2  \\
 3   & 75.3 & 89.7  \\
 4   & 80.6 & 95.9  \\
 \hline
\end{tabular}
\end{center} 
\label{table:eff}  
\end{table}

\subsection{Performance in the beam}

The detection efficiency was investigated further in the realistic experimental conditions with a $K^+$ beam. Charged particles were selected using using C2,C3 and C4 chambers, and the consistency of reconstructed tracks with the kaon decay point in the target was checked.  The detection efficiency 94~\% and 92~\% was obtained for the $R$-helicity and $L$-helicity, respectively, in the preliminary analysis.


More detailed performance of the SFT in the beam will be reported  in the future.


\section{Summary}

The SFT was  successfully constructed with the required position alignment for 1~mm fibers and  $z$ resolution sufficient for the E36 experiment. A method of assembling a spiral scintillating fiber detector by winding fiber ribbons  was established. The average detection efficiency was measured in the cosmic ray test to be nearly 80~\% per layer. The main contribution to the inefficiency  is attributed to the inactive material between the fibers (air gap and cladding). A staggered array of two layers removes this kind of inefficiency. The detection efficiency using four layers is, thus, sufficiently high. The E36 experiment equipped with the SFT  as the decay vertex detector has accumulated physics data in 2015 and the detailed data analysis is now underway. 


\section{Acknowledgments}
We are grateful to all TREK/E36 collaboration members for the valuable cooperation and detailed discussions regarding SFT. We deeply appreciate the support and encouragement of J-PARC and IPNS, KEK. This work was supported in part by the Natural Sciences and Engineering Research Council of Canada, JSPS KAKENHI grant 26287054 in Japan and the Russian Science Foundation Grant No.14-12-00560. 


\end{document}